\documentclass[12pt]{article}
\usepackage{amssym}
\def\Z{\mbox{$\Bbb Z$}}
\def\C{\mbox{$\Bbb C$}}
\def\case#1#2{{\textstyle{#1\over #2}}}
\def\ap{a^{\dagger}}
\def\bbeta{\bar{\beta}}
\def\bp{b^{\dagger}}

\title{
Spectrum generating algebra of the $C_{\lambda}$-extended oscillator and
multiphoton coherent states}

\author{C.\ Quesne \thanks{Directeur de recherches FNRS; E-mail:
cquesne@ulb.ac.be} \\
{\small \sl Physique Nucl\'eaire Th\'eorique et Physique
Math\'ematique,  Universit\'e Libre de Bruxelles,} \\ {\small \sl Campus de la
Plaine CP229, Boulevard~du Triomphe, B-1050 Brussels, Belgium}}
\date{ }
\begin{document}
\baselineskip=22pt plus 1pt minus 1pt
\maketitle

\begin{abstract}
The $C_{\lambda}$-extended oscillator spectrum generating algebra is shown to be
a $C_{\lambda}$-extended $(\lambda-1)$th-degree polynomial deformation of
su(1,1). Its coherent states are constructed. Their statistical and squeezing
properties are studied in detail. Such states include both some Barut-Girardello
and the standard $\lambda$-photon coherent states as special cases.
\end{abstract}

\vspace{0.5cm}






\newpage
%
%
\section{Introduction}

Coherent states~(CS) of the harmonic oscillator~\cite{glauber}, as well as
generalized CS associated with various algebras~\cite{perelomov}, have found
considerable applications in quantum optics. The former, defined as the
eigenstates
of the annihilation operator $a$, have properties similar to those of the
classical
radiation field. The latter, on the contrary, may exhibit some nonclassical
properties, such as photon antibunching~\cite{kimble} or sub-Poissonian photon
statistics~\cite{short}, and squeezing~\cite{slusher, hong}, which have
given rise
to an ever-increasing interest during the last few years.\par
%
%
As examples of CS with nonclassical properties, we may quote the eigenstates of
$a^2$~\cite{hillery}, which were introduced as even and odd CS or cat
states~\cite{dodonov}, and are a special case of generalized CS associated
with the
Lie algebra su(1,1)~\cite{perelomov,barut}. We may also mention the
eigenstates of
$a^{\lambda}$ ($\lambda>2$) or kitten states~\cite {buzek}, which may be
generated in $\lambda$-photon processes. Many alternative multiphoton CS have
been constructed and studied (see e.g.~\cite{dariano}).\par
%
%
Recently there has also been much interest in the study of nonlinear CS,
defined as
the eigenstates of the annihilation operator of a deformed oscillator (or
$f$-oscillator)~\cite{solomon,katriel,shanta,matos,manko}. It has indeed been
shown~\cite{matos} that for a particular class of nonlinearities they are
useful in
the description of a trapped ion and that they have strong nonclassical
properties.
Subsequently, even and odd nonlinear CS~\cite{mancini}, as well as the
eigenstates
of an arbitrary power of the $f$-oscillator annihilation
operator~\cite{liu}, have
also been considered in connection with nonclassical effects.\par
%
%
The purpose of the present Letter is to construct and study the nonclassical
properties of some multiphoton CS, which may be associated with the recently
introduced $C_{\lambda}$-extended oscillator~\cite{cq98}. The latter, which has
proved very useful in the context of supersymmetric quantum mechanics and some
of its variants~\cite{cq98,cq99}, may be considered as a deformed
oscillator with
a \Z$_{\lambda}$-graded Fock space. Hence, its CS will be a special case of the
nonlinear CS of Ref.~\cite{liu}. However, their connection with the
$C_{\lambda}$-extended oscillator spectrum generating algebra to be
determined in
the first part of this Letter will endow them with some extra properties. As a
consequence, they will satisfy Klauder's minimal set of conditions for
generalized
CS~\cite{klauder}, which is not the case in general for all the states of
Ref.~\cite{liu}.\par
%
%
\section{\boldmath The $C_{\lambda}$-extended oscillator and its spectrum
generating algebra}

The $C_{\lambda}$-extended oscillator Hamiltonian is defined (in units wherein
$\hbar \omega = 1$) by~\cite{cq98}
\begin{equation}
  H_0 = \case{1}{2} \left\{a, \ap\right\},
\end{equation}
where the creation and annihilation operators $\ap$, $a$ satisfy the relations
\begin{equation}
  \left[N, \ap\right] = \ap, \qquad \left[a, \ap\right] = I +
\sum_{\mu=0}^{\lambda-1}
  \alpha_{\mu} P_{\mu}, \qquad \ap P_{\mu} = P_{\mu+1} \ap, \label{eq:alg-def}
\end{equation}
together with their Hermitian conjugates. Here $N = N^{\dagger}$ is the number
operator, $\alpha_{\mu}$ are some real parameters subject to the conditions
$\sum_{\mu=0}^{\lambda-1} \alpha_{\mu} = 0$ and
\begin{equation}
  \sum_{\nu=0}^{\mu-1} \alpha_{\nu} > - \mu, \qquad \mu = 1, 2, \ldots,
\lambda-1,
  \label{eq:cond}
\end{equation}
and the operators $P_{\mu} = \lambda^{-1} \sum_{\nu=0}^{\lambda-1} \exp[2\pi
{\rm i} \nu (N-\mu)/\lambda]$, which are linear combinations of the
operators of a
cyclic group $C_{\lambda}$, project on the subspaces ${\cal F}_{\mu} \equiv
\{\, |k\lambda + \mu\rangle \mid k = 0, 1, \ldots\,\}$ of the
\Z$_{\lambda}$-graded Fock space ${\cal F} = \sum_{\mu=0}^{\lambda-1} \oplus
{\cal F}_{\mu}$. Throughout this paper, we use the convention $P_{\mu'} =
P_{\mu}$
if $\mu' - \mu = 0\, {\rm mod}\, \lambda$ (and similarly for other operators or
parameters labelled by $\mu$, $\mu'$). The operators $N$, $\ap$, $a$ are
related to each other through the structure function $F(N)$, which is a
fundamental concept of deformed oscillators~\cite{solomon,daska}: $\ap a =
F(N)$,
$a \ap = F(N+1)$. In the present case, $F(N)$ is given by $F(N) = N +
\sum_{\mu=0}^{\lambda-1} \beta_{\mu} P_{\mu}$, where $\beta_{\mu} \equiv
\sum_{\nu=0}^{\mu-1} \alpha_{\nu}$ (with $\beta_0 \equiv 0$)~\cite{cq98}.\par
%
%
The Fock space basis states $|n\rangle = |k\lambda + \mu\rangle$, $k = 0$,
1,~\ldots, $\mu=0$, 1,~\ldots, $\lambda-1$, are given by $|n\rangle = {\cal
N}_n^{-1/2} \left(\ap\right)^n |0\rangle$, where $|0\rangle$ is a vacuum
state (i.e.,
$a |0\rangle = 0$), and ${\cal N}_n$ is some normalization coefficient~\cite{cq98}.
The number, creation, and annihilation operators act on $|n\rangle$ as
\begin{equation}
  N |n\rangle = n |n\rangle, \qquad \ap |n\rangle = \sqrt{F(n+1)}\,
|n+1\rangle, \qquad
  a |n\rangle = \sqrt{F(n)}\, |n-1\rangle. \label{eq:op-action}
\end{equation}
From the restriction~(\ref{eq:cond}) on the parameters, it follows that $F(\mu)
= \beta_{\mu} + \mu > 0$, so that all the states $|n\rangle$ are well
defined.\par
%
%
The eigenstates of $H_0$ are the states $|n\rangle = |k\lambda + \mu\rangle$ and
their eigenvalues are given by $E_{k\lambda+\mu} = k\lambda + \mu +
\gamma_{\mu} + \frac{1}{2}$, where $\gamma_{\mu} \equiv \frac{1}{2}
\left(\beta_{\mu} + \beta_{\mu+1}\right)$. In each ${\cal F}_{\mu}$ subspace of
$\cal F$, the spectrum of $H_0$ is harmonic, but the $\lambda$ infinite sets of
equally spaced energy levels, corresponding to $\mu=0$, 1,~\ldots, $\lambda-1$,
are shifted with respect to each other by some amounts depending upon the
parameters $\alpha_0$, $\alpha_1$,~\ldots, $\alpha_{\lambda-1}$.\par
%
%
{}For vanishing parameters $\alpha_{\mu}$, the $C_{\lambda}$-extended oscillator
reduces to the standard harmonic oscillator. In such a case, $\left[a,
\ap\right] =
I$, $\beta_{\mu} = \gamma_{\mu} = 0$, $F(N) = N$, and $E_n = n +
\frac{1}{2}$. The
operators
\begin{equation}
  J_+ = \case{1}{2} \left(\ap\right)^2, \qquad J_- = \case{1}{2} a^2, \qquad
  J_0 = \case{1}{2} H_0 = \case{1}{4} \left\{a, \ap\right\}  \label{eq:su-gen}
\end{equation}
are known to generate the whole spectrum from the zero- and one-quantum
states~\cite{moshinsky}. They satisfy the commutation relations
\begin{equation}
  \left[J_0, J_{\pm}\right] = \pm J_{\pm}, \qquad \left[J_+, J_-\right] = -
2J_0,
  \label{eq:su-com}
\end{equation}
and the Hermiticity properties $J_0^{\dagger} = J_0$, $J_{\pm}^{\dagger} =
J_{\mp}$, characteristic of the Lie algebra su(1,1). The Casimir operator
\begin{equation}
  C = J_+ J_- - J_0 (J_0-1) = J_- J_+ - J_0 (J_0+1), \label{eq:su-Cas}
\end{equation}
which commutes with $J_0$, $J_+$, $J_-$, has the same eigenvalue $c = 3/16$ in
the two su(1,1) unitary irreducible representations (unirreps) corresponding to
even and odd states. The latter are distinguished by the lowest $J_0$
eigenvalue,
equal to $1/4$ and $3/4$, respectively.\par
%
%
{}For nonvanishing parameters and $\lambda=2$, the $C_2$-extended oscillator is
equivalent to the Calogero-Vasiliev oscillator, which provides an algebraic
formulation of the two-particle Calogero problem (see~\cite{brze} and references
quoted therein), and an alternative description of
parabosons~\cite{chaturvedi}. In
such a case, $\left[a, \ap\right] = I + \alpha_0 (-1)^N$, $\alpha_0 = -
\alpha_1 =
\beta_1 = 2 \gamma_0 = 2 \gamma_1$, $F(N) = N + \alpha_0 [1 - (-1)^N]/2$, and
$E_n = n + (\alpha_0+1)/2$. The spectrum is that of a shifted oscillator and the
operators~(\ref{eq:su-gen}) still form an su(1,1) spectrum generating
algebra~\cite{brze,mukunda}. The Casimir operator~(\ref{eq:su-Cas}) has now
distinct eigenvalues $c_{\mu} = (1 + \alpha_{\mu}) (3 - \alpha_{\mu})/16$
for even
($\mu=0$) and odd ($\mu=1$) states, and the lowest $J_0$ eigenvalue is
$(1+\alpha_0)/4$ and $(3+\alpha_0)/4$, respectively.\par
%
%
When going to $\lambda$ values greater than two, the operators~(\ref{eq:su-gen})
are replaced by the operators
\begin{equation}
  J_+ = \frac{1}{\lambda} \left(\ap\right)^{\lambda}, \qquad J_- =
  \frac{1}{\lambda} a^{\lambda}, \qquad J_0 = \frac{1}{\lambda} H_0 =
  \frac{1}{2\lambda} \left\{a, \ap\right\}, \label{eq:defsu-gen}
\end{equation}
which connect among themselves all the equally spaced levels characterized by a
given $\mu$ value. By using~(\ref{eq:alg-def}) or~(\ref{eq:op-action}), it
is easy to
show that they satisfy the commutation relations
\begin{equation}
  [J_0, J_{\pm}] = \pm J_{\pm}, \qquad [J_+, J_-] = f(J_0, P_{\mu}),
  \label{eq:defsu-com}
\end{equation}
where
\begin{eqnarray}
  && f(J_0, P_{\mu}) = - \frac{1}{\lambda} \Biggl\{\prod_{l=0}^{\lambda-2}
\left(
         \lambda J_0 + \case{1}{2} \sum_{\mu} \left(2l + 1 + \alpha_{\mu} + 2
         \sum_{m=1}^l \alpha_{\mu+m}\right) P_{\mu}\right) \nonumber \\
  && \mbox{} + \sum_{i=1}^{\lambda-1} \left(\lambda J_0 - \case{1}{2}
         \sum_{\mu} (1 + \alpha_{\mu}) P_{\mu}\right) \left[\prod_{j=1}^{i-1}
         \left(\lambda J_0 + \case{1}{2} \sum_{\mu} \left(- 2j - 1 +
\alpha_{\mu} + 2
         \sum_{k=1}^{\lambda-j-1} \alpha_{\mu+k}\right) P_{\mu}\right)\right]
         \nonumber \\
  && \mbox{} \times \left[\prod_{l=0}^{\lambda-i-2} \left(\lambda J_0 +
         \case{1}{2} \sum_{\mu} \left(2l + 1 + \alpha_{\mu} + 2 \sum_{m=1}^l
         \alpha_{\mu+m}\right) P_{\mu}\right)\right]\Biggr\} \label{eq:f}
\end{eqnarray}
is a ($\lambda-1$)th-degree polynomial in $J_0$ with $P_{\mu}$-dependent
coefficients, $f(J_0, P_{\mu}) = \sum_{i=0}^{\lambda-1} s_i(P_{\mu}) J_0^i$. The
definition of the spectrum generating algebra is completed by the commutation
relations
\begin{equation}
  [J_0, P_{\mu}] = [J_+, P_{\mu}] = [J_-, P_{\mu}] = 0.
\end{equation}
Since the $P_{\mu}$'s are linear combinations of $C_{\lambda}$ operators, we
conclude that the algebra is a $C_{\lambda}$-extended polynomial deformation of
su(1,1): in each ${\cal F}_{\mu}$ subspace, it reduces to a standard polynomial
deformation of su(1,1)~\cite{poly}.\par
%
%
Its Casimir operator can be written as
\begin{equation}
  C = J_- J_+ + h(J_0, P_{\mu}) = J_+ J_- + h(J_0, P_{\mu}) - f(J_0, P_{\mu}),
  \label{eq:defsu-Cas}
\end{equation}
where
\begin{eqnarray}
  h(J_0, P_{\mu}) & = & \frac{1}{\lambda^2} \Biggl\{- \left[\lambda J_0 +
         \case{1}{2} \sum_{\mu} (2\lambda - 1 - \alpha_{\mu}) P_{\mu}\right]
         \nonumber \\
  && \mbox{} \times \prod_{k=1}^{\lambda-1} \left[\lambda J_0 +
         \case{1}{2} \sum_{\mu} \left(2k - 1 + \alpha_{\mu} + 2 \sum_{l=1}^{k-1}
         \alpha_{\mu+l}\right) P_{\mu}\right] \nonumber \\
  && \mbox{} + \frac{1}{2^{\lambda}} \sum_{\mu} \left[(2\lambda - 1 -
         \alpha_{\mu}) \prod_{k=1}^{\lambda-1} \left(2k - 1 + \alpha_{\mu} + 2
         \sum_{l=1}^{k-1} \alpha_{\mu+l}\right) P_{\mu}\right]\Biggr\}
         \label{eq:h}
\end{eqnarray}
is a $\lambda$th-degree polynomial in $J_0$ with $P_{\mu}$-dependent
coefficients, $h(J_0, P_{\mu}) = \sum_{i=0}^{\lambda} t_i(P_{\mu}) J_0^i$.\par
%
%
Each ${\cal F}_{\mu}$ subspace is the carrier space of a unirrep
characterized by
the eigenvalue
\begin{equation}
  c_{\mu} = \frac{1}{\lambda^2 2^{\lambda}} (2\lambda - 1 - \alpha_{\mu})
  \prod_{k=1}^{\lambda-1} \left(2k - 1 + \alpha_{\mu} + 2 \sum_{l=1}^{k-1}
  \alpha_{\mu+l}\right)
\end{equation}
of $C$, and by the lowest eigenvalue $\left(\mu + \gamma_{\mu} +
\case{1}{2}\right)/{\lambda}$ of $J_0$.\par
%
%
{}For $\lambda=2$, equations~(\ref{eq:defsu-com}), (\ref{eq:f}),
(\ref{eq:defsu-Cas}), and~(\ref{eq:h}) reduce to equations~(\ref{eq:su-com})
and~(\ref{eq:su-Cas}), as it should be. Nonlinearities make their appearance for
$\lambda=3$, for which
\begin{eqnarray}
  f(J_0, P_{\mu}) & = & - 9 J_0^2 - J_0 \sum_{\mu} (\alpha_{\mu} +
        2\alpha_{\mu+1}) P_{\mu} - \case{1}{12} \sum_{\mu} (1 + \alpha_{\mu})
        (5 - \alpha_{\mu}) P_{\mu}, \nonumber \\
  h(J_0, P_{\mu}) & = & - J_0 \biggl[3 J_0^2 + \case{1}{2} J_0 \sum_{\mu}
        (9 + \alpha_{\mu} + 2\alpha_{\mu+1}) P_{\mu} + \case{1}{12} \sum_{\mu}
        \bigl(23 + 10 \alpha_{\mu} \nonumber \\
  && \mbox{} + 12 \alpha_{\mu+1} - \alpha_{\mu}^2\bigr) P_{\mu}\biggr],
        \nonumber  \\
  c_{\mu} & = & \case{1}{72} (1 + \alpha_{\mu}) (5 - \alpha_{\mu}) (3 +
        \alpha_{\mu} + 2 \alpha_{\mu+1}),
\end{eqnarray}
and the lowest $J_0$ eigenvalues are $(1 + \alpha_0)/6$, $(3 - \alpha_1 -
2\alpha_2)/6$, $(5 - \alpha_2)/6$ for $\mu=0$, 1, 2, respectively.\par
%
%
As a by-product of our analysis, it is also worth mentioning that for
$\alpha_{\mu}
= 0$, the operators~(\ref{eq:defsu-gen}) close a polynomial deformation of
su(1,1),
characterized by equations~(\ref{eq:defsu-com}) and~(\ref{eq:defsu-Cas}), where
$f(J_0, P_{\mu})$ and $h(J_0, P_{\mu})$ are replaced by
\begin{eqnarray}
  f(J_0) & = & \frac{1}{\lambda^2} \sum_{j=0}^{\lambda} \left[1 -
(-1)^{\lambda-j}
          \right] (\lambda J_0)^j \sum_{i=j}^{\lambda}
\left({}^{\displaystyle i}
          _{\displaystyle j}\right) \left(- \frac{1}{2}\right)^{i-j}
S^{(i)}_{\lambda},
          \nonumber \\
  h(J_0) & = & - \frac{1}{\lambda^2} \sum_{j=1}^{\lambda} (\lambda J_0)^j
          \sum_{i=j}^{\lambda} (-1)^{\lambda-i} \left({}^{\displaystyle i}
          _{\displaystyle j}\right) \left(\frac{1}{2}\right)^{i-j}
S^{(i)}_{\lambda},
\end{eqnarray}
and $\left({}^i_j\right)$, $S^{(i)}_{\lambda}$ denote a binomial
coefficient and a
Stirling number of the first kind, respectively. Note that the function
$f(J_0)$ has
a given parity, opposite to that of $\lambda$, and that the eigenvalues of
$C$ do
not depend on $\mu$ and are given by $c = (2\lambda - 1)!!\,/\left(\lambda^2
2^{\lambda}\right)$.\par
%
%
\section{\boldmath Coherent states associated with the $C_{\lambda}$-extended
oscillator spectrum generating algebra}

As mentioned in the previous section, the $C_2$-extended oscillator spectrum
generating algebra is the Lie algebra su(1,1), with which one can associate
various
types of generalized CS (see e.g.~\cite{perelomov,barut}). Of special
relevance in
quantum optics are the Barut-Girardello CS~\cite{barut}, which are the
eigenstates
of the generator $J_-$, defined in~(\ref{eq:su-gen}).\par
%
%
{}For the $C_{\lambda}$-extended oscillator with $\lambda > 2$, it
therefore seems
appropriate to consider as CS the eigenstates $|z; \mu\rangle$ of the operator
$J_-$, defined in~(\ref{eq:defsu-gen}),
\begin{equation}
  J_- |z; \mu\rangle = z |z; \mu\rangle, \qquad z \in \C, \qquad \mu=0, 1,
\ldots,
  \lambda-1.  \label{eq:CS-def}
\end{equation}
Here $\mu$ distinguishes between the $\lambda$ independent (and orthogonal)
solutions of equation~(\ref{eq:CS-def}), belonging to the various subspaces
${\cal
F}_{\mu}$.\par
%
%
The CS $|z; \mu\rangle$ may be considered as special cases of the nonlinear
CS of
Ref.~\cite{liu}, since $J_-$ may be written in terms of the creation and
annihilation
operators $\bp$, $b$ of a standard harmonic oscillator as
\begin{equation}
  J_- = (b f(N_b))^{\lambda}, \qquad N_b \equiv \bp b = N, \qquad f(N_b) =
  \lambda^{-1/\lambda} \left[\frac{F(N_b)}{N_b}\right]^{1/2}.  \label{eq:liu}
\end{equation}
More interesting for our purposes, however, is the similarity existing
between $|z;
\mu\rangle$ and some CS of nonlinear algebras~\cite{junker}, when disregarding
the discrete label $\mu$ occurring in the former.\par
%
%
By using (\ref{eq:op-action}) and~(\ref{eq:defsu-gen}), it is easy to
construct the
CS $|z; \mu\rangle$ in terms of the basis states $|k \lambda + \mu\rangle$ of
${\cal F}_{\mu}$. The result reads
\begin{equation}
  |z; \mu\rangle = \left[N_{\mu}(|z|)\right]^{-1/2} \sum_{k=0}^{\infty}
  \frac{\left(z/\lambda^{(\lambda-2)/2}\right)^k}{\left[k!\,
  \left(\prod_{\nu=1}^{\mu} (\bbeta_{\nu}+1)_k\right)
  \left(\prod_{\nu'=\mu+1}^{\lambda-1} (\bbeta_{\nu'})_k\right)\right]^{1/2}}
  |k \lambda + \mu\rangle, \label{eq:CS-exp}
\end{equation}
where $\bbeta_{\mu} \equiv (\beta_{\mu} + \mu)/\lambda$, $(a)_k$ denotes
Pochhammer's symbol, and the normalization factor $N_{\mu}(|z|)$ can be expressed
in terms of a generalized hypergeometric function,
\begin{equation}
  N_{\mu}(|z|) = {}_0F_{\lambda-1} \left(\bbeta_1+1, \ldots, \bbeta_{\mu}+1,
  \bbeta_{\mu+1}, \ldots, \bbeta_{\lambda-1}; y\right), \qquad y \equiv
  |z|^2/\lambda^{\lambda-2}.
\end{equation}
There exists an alternative form of~(\ref{eq:CS-exp}) in terms of the generator
$J_+$ of the spectrum generating algebra,
\begin{equation}
  |z; \mu\rangle = \left[N_{\mu}(|z|)\right]^{-1/2} {}_0F_{\lambda-1}
  \left(\bbeta_1+1, \ldots, \bbeta_{\mu}+1, \bbeta_{\mu+1}, \ldots,
  \bbeta_{\lambda-1}; z J_+/\lambda^{\lambda-2}\right) |\mu\rangle.
\end{equation}
\par
%
%
The set of CS $\left\{\, |z, \mu\rangle \mid \mu=0, 1, \ldots, \lambda-1\,
\right\}$
satisfies a unity resolution relation, which can be written as
\begin{equation}
  \sum_{\mu} \int d\rho_{\mu}\left(z, z^*\right) |z; \mu\rangle \langle z;
\mu| = I,
  \label{eq:unity}
\end{equation}
where $d\rho_{\mu}\left(z, z^*\right)$ is a positive measure. Making the polar
decomposition $z = |z| \exp(i\phi)$ and the ansatz $d\rho_{\mu}\left(z,
z^*\right) =
{}_0F_{\lambda-1} \left(\bbeta_1+1, \ldots, \bbeta_{\mu}+1, \bbeta_{\mu+1},
\ldots, \bbeta_{\lambda-1}; y\right) h_{\mu}(y) |z| d|z| d\phi$, where
$h_{\mu}(y)$
is a yet unknown density on the positive half-line, we find that
equation~(\ref{eq:unity}) reduces to the relations
\begin{equation}
  \int_0^{\infty} dy\, y^k h_{\mu}(y) = \frac{k!}{\pi \lambda^{\lambda-2}}
  \left(\prod_{\nu=1}^{\mu} (\bbeta_{\nu}+1)_k\right)
  \left(\prod_{\nu'=\mu+1}^{\lambda-1} (\bbeta_{\nu'})_k\right),
\label{eq:mellin}
\end{equation}
where $k=0$, 1,~\ldots, and $\mu=0$, 1,~\ldots, $\lambda-1$. Hence $h_{\mu}(y)$
is the inverse Mellin transform of the right-hand side of~(\ref{eq:mellin})
and is
proportional to a Meijer $G$-function~\cite{erdelyi54}:
\begin{equation}
  h_{\mu}(y) = \frac{G^{\lambda 0}_{0 \lambda} \left(y \mid 0, \bbeta_1, \ldots,
  \bbeta_{\mu}, \bbeta_{\mu+1}-1, \ldots, \bbeta_{\lambda-1}-1\right)}
  {\pi \lambda^{\lambda-2} \left(\prod_{\nu=1}^{\mu} \Gamma
  (\bbeta_{\nu}+1)\right) \left(\prod_{\nu'=\mu+1}^{\lambda-1} \Gamma
  (\bbeta_{\nu'})\right)}.  \label{eq:measure}
\end{equation}
\par
%
%
There are two main consequences arising from the latter result. First, we can
express any CS in terms of the others corresponding to the same $\mu$ value:
\begin{equation}
  |z; \mu\rangle = \int d\rho_{\mu}\left(z', z^{\prime*}\right) |z'; \mu\rangle
  \langle z', \mu | z; \mu\rangle.
\end{equation}
The reproducing kernel $\langle z', \mu' | z; \mu\rangle$ can be evaluated
from~(\ref{eq:CS-exp}) and is given by
\begin{eqnarray}
  \langle z', \mu' | z; \mu\rangle & = & \delta_{\mu',\mu} \left[N_{\mu}(|z|)
         N_{\mu}(|z'|)\right]^{-1/2} \nonumber \\
  && \mbox{} \times {}_0F_{\lambda-1} \left(\bbeta_1+1, \ldots, \bbeta_{\mu}+1,
         \bbeta_{\mu+1}, \ldots, \bbeta_{\lambda-1}; z^{\prime*} z
         /\lambda^{\lambda-2}\right).
\end{eqnarray}
Second, an arbitrary element $|\psi\rangle$ of the Fock space $\cal F$ can be
written in terms of the CS:
\begin{equation}
  |\psi\rangle = \sum_{\mu} \int d\rho_{\mu}\left(z, z^*\right)
  \tilde{\psi}_{\mu}\left(z, z^*\right) | z; \mu\rangle,
\end{equation}
where
\begin{equation}
  \tilde{\psi}_{\mu}\left(z, z^*\right) = \left[N_{\mu}(|z|)\right]^{-1/2}
  \sum_{k=0}^{\infty}
\frac{\left(z^*/\lambda^{(\lambda-2)/2}\right)^k}{\left[k!\,
  \left(\prod_{\nu=1}^{\mu} (\bbeta_{\nu}+1)_k\right)
  \left(\prod_{\nu'=\mu+1}^{\lambda-1} (\bbeta_{\nu'})_k\right)\right]^{1/2}}\,
  \langle k \lambda + \mu | \psi\rangle.
\end{equation}
All these properties show that the CS form an overcomplete basis of $\cal
F$.\par
%
%
{}From the previous results, it follows that the CS, defined in
equation~(\ref{eq:CS-def}), satisfy Klauder's minimal set of conditions for
generalized CS~\cite{klauder}: they are normalizable, continuous in the
label $z$,
and they allow a resolution of unity. It is also worth mentioning that the other
discrete label $\mu$ is analogous to the vector components of vector (or
partially)
CS~\cite{deenen}.\par
%
%
We conclude the present section by presenting two important special cases of our
CS. The first one corresponds to $\lambda=2$ and $\alpha_{\mu} \ne 0$. In such a
case, the generalized hypergeometric function ${}_0F_1$ and the Meijer
$G$-function $G^{20}_{02}$ reduce to modified Bessel functions $I_{\nu}$ and
$K_{\nu}$ for some appropriate $\nu$ value~\cite{erdelyi53}, respectively,
so that
\begin{eqnarray}
  |z; \mu\rangle & = & \left(\frac{|z|^{(\alpha_0-1+2\mu)/2}}
         {I_{(\alpha_0-1+2\mu)/2}(2|z|)}\right)^{1/2} \sum_{k=0}^{\infty}
\frac{z^k}
         {[k!\, \Gamma((\alpha_0+1+2\mu+2k)/2)]^{1/2}}\, |2k + \mu\rangle,
         \nonumber \\
  d\rho_{\mu}\left(z, z^*\right) & = & 2 \pi^{-1}
         I_{(\alpha_0-1+2\mu)/2}(2|z|) K_{(\alpha_0-1+2\mu)/2}(2|z|)\, |z| d|z|
         d\phi,
\end{eqnarray}
where $\mu=0$, 1. This gives back Barut-Girardello results for the su(1,1)
unirrep
characterized by the lowest $J_0$ eigenvalue $(\alpha_0 + 1 + 2\mu)/4$.\par
%
%
The second case corresponds to an arbitrary value of $\lambda$ and $\alpha_{\mu}
= 0$. By taking into account that now $\bbeta_{\mu} = \mu/\lambda$, the CS given
in equation~(\ref{eq:CS-exp}) reduce to the standard $\lambda$-photon
CS~\cite{buzek},
\begin{equation}
  |z; \mu\rangle = [N_{\mu}(|z|)]^{-1/2} \sum_{k=0}^{\infty}
  \left(\frac{\mu!}{(k\lambda+\mu)!}\right)^{1/2} (\lambda z)^k |k\lambda
  + \mu\rangle.  \label{eq:buzek-CS}
\end{equation}
From~(\ref{eq:unity}), it follows that such CS satisfy a unity resolution
relation,
which, as far as the author knows, is a new result. To find the special
form taken
by $h_{\mu}(y)$ in~(\ref{eq:measure}), the easiest thing is to go back to
equation~(\ref{eq:mellin}) and to rewrite it as
\begin{equation}
  \int_0^{\infty} dy\, y^k h_{\mu}(y) = \lambda^{2 - \lambda(k+1)} \left(\pi
  \mu!\right)^{-1} \Gamma(k\lambda + \mu + 1),
\end{equation}
by using Gauss' multiplication formula. Then an inverse Mellin
transform~\cite{erdelyi54} directly leads to
\begin{equation}
  h_{\mu}(y) = \lambda^{\mu-\lambda+2} \left(\pi \mu!\right)^{-1}
  y^{(\mu-\lambda+1)/\lambda} \exp\left(- \lambda y^{1/\lambda}\right).
  \label{eq:buzek-weight}
\end{equation}
\par
%
%
It should be noted that the states~(\ref{eq:buzek-CS}) provide us with a simple
example of Mittag-Leffler CS~\cite{sixdeniers}, since $N_{\mu}(|z|)$ and
$|z;\mu\rangle$ can be written as
\begin{eqnarray}
  N_{\mu}(|z|) & = & \mu!\, E_{\lambda,\mu+1}\left(\lambda^2 |z|^2\right), \\
  |z;\mu\rangle & = & \left(\frac{\mu!}{E_{\lambda,\mu+1}\left(\lambda^2
         |z|^2\right)}\right)^{1/2} E_{\lambda,\mu+1}\left(\lambda^2 z
J_+\right)
         |\mu\rangle,
\end{eqnarray}
where $E_{\alpha,\beta}(x) \equiv \sum_{k=0}^{\infty} x^k/\Gamma(\alpha k +
\beta)$ is a generalized Mittag-Leffler function~\cite{erdelyi53}. The weight
function~(\ref{eq:buzek-weight}) agrees with the principal solution of
Ref.~\cite{sixdeniers}, obtained for arbitrary positive values of $\alpha$,
$\beta$.
In addition, our results show that for $\alpha = \lambda$, $\beta = \mu +
1$, the
deformed boson operators $\hat{b}^{\dagger}_{\alpha,\beta}$,
$\hat{b}_{\alpha,\beta}$, and $\hat{N}$ of Ref.~\cite{sixdeniers} can be
realized as
$\left(\bp\right)^{\lambda}$, $b^{\lambda}$, and $(N_b - \mu)/\lambda$,
respectively, where $\bp$, $b$, and $N_b$ are the standard boson operators
considered in~(\ref{eq:liu}). Here the roles of $z$ and of the vacuum state
$|0\rangle$ are played by $\lambda z$ and $|\mu\rangle$, respectively.\par
%
%
\section{Nonclassical properties of coherent states}

In quantum optics, the properties of the CS $|z;\mu\rangle$ may be analyzed
in two
different ways. In both approaches, they are considered as exotic states
defined in
terms of the deformed operators $\ap$, $a$, but in the first one considers
``real''
photons, described by the operators $\bp$, $b$ of~(\ref{eq:liu}) satisfying the
canonical commutation relation, while in the second one considers ``dressed''
photons, described by the operators $\ap$, $a$ satisfying a more general
commutation relation. Such generalized photons may be invoked in
phenomenological models explaining some non-intuitive observable phenomena.\par
%
%
We use here the latter approach, leaving the former for future work. It
should be
stressed that this choice only affects the squeezing properties (to be
studied in
Subsec.~4.2) through the definition of the quadratures. On the contrary,
since $N$
and $N_b$ coincide (see~(\ref{eq:liu})), the deformed photon statistics (to be
studied in Subsec.~4.1) is actually the same as the photon statistics,
which would
result from the other approach.\par
%
%
\subsection{Photon statistics}

A convenient measure of the deviation of the photon number statistics from the
Poisson distribution is the Mandel parameter
\begin{equation}
  Q = \frac{(\Delta N)^2 - \langle N \rangle}{\langle N \rangle},
\end{equation}
which vanishes for the Poisson distribution. It is positive or negative
according to
whether the distribution is super-Poissonian (bunching effect) or sub-Poissonian
(antibunching effect).\par
%
%
{}From~(\ref{eq:op-action}) and~(\ref{eq:CS-exp}), we obtain
\begin{eqnarray}
  Q & = & \lambda \left[1 - \bbeta_{\lambda-1} - \left(\prod_{\nu=1}^{\lambda-1}
         \bbeta_{\nu}\right)^{-1} y\, \Phi^{\lambda-1}_0(y) + \bbeta_{\lambda-1}
         \Phi^{\lambda-2}_{\lambda-1}(y)\right] - 1 \qquad \mbox{\rm if\ }
\mu=0,
         \nonumber \\
  & = & \left[\lambda^{-1} - \bbeta_1 + \bbeta_1 \Phi^0_1(y)\right]^{-1}
         \Biggl\{\left(\bbeta_1 - \lambda^{-1}\right) \left[1 + \lambda \bbeta_1
         \Phi^0_1(y)\right] - \lambda \bbeta_1^2 \left[\Phi^0_1(y)\right]^2
         \nonumber \\
  && \mbox{} + \lambda \left(\prod_{\nu=2}^{\lambda-1} \bbeta_{\nu}\right)^{-1}
         y\, \Phi^{\lambda-1}_1(y)\Biggr\} \qquad \mbox{\rm if\ } \mu=1,
         \nonumber \\
  & = & \left[\mu \lambda^{-1} - \bbeta_{\mu} + \bbeta_{\mu}
         \Phi^{\mu-1}_{\mu}(y)\right]^{-1} \Biggl\{\bbeta_{\mu} - \mu
\lambda^{-1}
         + \lambda \bbeta_{\mu} \left(\bbeta_{\mu} - \bbeta_{\mu-1} -
\lambda^{-1}
         \right) \Phi^{\mu-1}_{\mu}(y) \nonumber \\
  && \mbox{} - \lambda \bbeta_{\mu}^2 \left[\Phi^{\mu-1}_{\mu}(y)\right]^2 +
         \lambda \bbeta_{\mu-1} \bbeta_{\mu} \Phi^{\mu-2}_{\mu}(y)\Biggr\}
\qquad
         \mbox{\rm if\ } \mu=2, 3, \ldots, \lambda-1,  \label{eq:Q}
\end{eqnarray}
where $y = |z|^2/\lambda^{\lambda-2}$ and $\Phi^{\mu'}_{\mu}(y) = N_{\mu'}(|z|)/
N_{\mu}(|z|)$.\par
%
%
Standard even (resp.\ odd) CS, corresponding to $\lambda=2$, $\alpha_0 =
\alpha_1
= 0$, $\mu=0$ (resp.\ $\mu=1$) are known to exhibit a bunching (resp.\
antibunching) effect. For the even (resp.\ odd) CS associated with the
Calogero-Vasiliev oscillator, i.e., for $\lambda=2$, $\alpha_0 = - \alpha_1
\ne 0$,
$\mu=0$ (resp.\ $\mu=1$), this trend is enhanced for positive (resp.\ negative)
values of $\alpha_0$. However, as shown in Fig.~1, for negative (resp.\
positive)
values of $\alpha_0$ and sufficiently high values of $|z|$, the opposite
trend can be
seen. In particular, for well-chosen values of $\alpha_0$, it is possible to get
antibunching for even CS over almost the whole $|z|$ range.\par
%
%
{}For higher values of $\lambda$, more or less similar results are obtained for
$\mu=0$, on one hand, and $\mu \ne 0$, on the other hand.
From~(\ref{eq:Q}), it is
straightforward to show that for any values of $\alpha_{\mu}$ and $|z|=0$, $Q =
\lambda-1$ if $\mu=0$, and $Q = -1$ if $\mu=1$, 2,~\ldots, $\lambda-1$. Hence
there is bunching (resp.\ antibunching) for $\mu=0$ (resp.\ $\mu \ne 0$) for
sufficiently small values of $|z|$. Fig.~2, corresponding to $\lambda=3$,
shows that
considering negative (resp.\ positive) values of $\alpha_0$ ($= \beta_1$) or/and
$\alpha_0 + \alpha_1$ ($= \beta_2$) allows one to reverse the trend and to get
antibunching (resp.\ bunching) for $\mu=0$ (resp.\ $\mu \ne 0$) for sufficiently
high values of $|z|$. Note however that the behaviour of $Q$ is more
complicated for
$\mu=1$ than for $\mu=0$ or~2.\par
%
%
\subsection{Squeezing effect}

Let us define the deformed electromagnetic field components $x$ and $p$ as
\begin{equation}
  x = \frac{1}{\sqrt{2}} \left(\ap + a\right), \qquad p = \frac{{\rm
i}}{\sqrt{2}}
  \left(\ap - a\right).
\end{equation}
In any state belonging to ${\cal F}_{\mu}$, their dispersions $\langle
(\Delta x)^2
\rangle$ and $\langle (\Delta p)^2 \rangle$, where $\Delta x \equiv x -
\langle x
\rangle$ and $\Delta p \equiv p - \langle p \rangle$, satisfy the uncertainty
relation
\begin{equation}
 \langle (\Delta x)^2 \rangle \langle (\Delta p)^2 \rangle \ge \frac{1}{4}
|\langle [x,
  p]\rangle|^2 = \frac{\lambda^2}{4} (\bbeta_{\mu+1} - \bbeta_{\mu})^2.
  \label{eq:UR}
\end{equation}
We note that the right-hand side of this inequality becomes smaller than the
conventional value 1/4 if $\alpha_0 < 0$ for $\mu=0$ or $-2 < \alpha_{\mu} < 0$
for $\mu=1$, 2, \ldots, or $\lambda-1$. In the latter case, it even vanishes for
$\alpha_{\mu} = -1$.\par
%
%
Here we are interested in the dispersions in the CS $|z;\mu\rangle$, which are
obtained as
\begin{equation}
  \langle (\Delta x)^2\rangle = \langle H_0 \rangle + \delta_{\lambda,2}
\left(z +
  z^*\right), \qquad \langle (\Delta p)^2\rangle = \langle H_0 \rangle -
  \delta_{\lambda,2} \left(z + z^*\right),  \label{eq:dispersions}
\end{equation}
where
\begin{eqnarray}
  \langle H_0 \rangle & = & \lambda \left[\case{1}{2} \bbeta_1 +
          \left(\prod_{\nu=1}^{\lambda-1} \bbeta_{\nu}\right)^{-1} y\,
          \Phi^{\lambda-1}_0(y)\right] \qquad \mbox{\rm if\ } \mu=0,
\nonumber \\
  & = & \lambda \left[\case{1}{2} (\bbeta_{\mu+1} - \bbeta_{\mu}) + \bbeta_{\mu}
          \Phi^{\mu-1}_{\mu}(y)\right] \qquad \mbox{\rm if\ } \mu = 1, 2,
\ldots,
          \lambda-1.  \label{eq:H_0}
\end{eqnarray}
\par
%
%
In ${\cal F}_{\mu}$, the role of the vacuum state is played by the number state
$|\mu\rangle = |0;\mu\rangle$, which is annihilated by $J_-$. The corresponding
dispersions are given by
\begin{equation}
  \langle (\Delta x)^2\rangle_0 = \langle (\Delta p)^2\rangle_0 =
\frac{\lambda}{2}
  (\bbeta_{\mu+1} + \bbeta_{\mu}) = \gamma_{\mu} + \mu + \frac{1}{2}.
  \label{eq:dispersions-vac}
\end{equation}
Comparing with the uncertainty relation~(\ref{eq:UR}), we conclude that the
state
$|\mu\rangle$ satisfies the minimum uncertainty property in ${\cal
F}_{\mu}$, i.e.,
gives rise to the equality in~(\ref{eq:UR}), only for $\mu=0$ because
$\bbeta_0 =
0$ and $\bbeta_{\mu} > 0$ for $\mu=1$, 2,~\ldots, $\lambda-1$. On the other
hand,
the dispersions in the vacuum are smaller than the conventional value 1/2 if
$\gamma_{\mu} < - \mu$. However, from condition~(\ref{eq:cond}) and the
definitions of $\beta_{\mu}$ and $\gamma_{\mu}$, it follows that
$\gamma_{\mu}$ is also restricted by the condition $\gamma_{\mu} > - \mu -
\frac{1}{2}$ if $\mu=0$, 1,~\ldots, $\lambda - 2$, or $\gamma_{\mu} > -
\mu/2$ if
$\mu = \lambda-1$. Since both types of conditions on $\gamma_{\mu}$ are
compatible only for $\mu=0$, 1,~\ldots, $\lambda-2$, we conclude that $\langle
(\Delta x)^2\rangle_0 = \langle (\Delta p)^2\rangle_0$ can be less than 1/2 for
such $\mu$ values only..\par
%
%
In the following, we shall restrict ourselves to $\mu=0$, for which the vacuum
state $|0\rangle$ satisfies the minimum uncertainty property. According to the
usual definition~\cite{solomon}, we say that the quadrature $x$ (resp.\ $p$) is
squeezed in $|z;0\rangle$ if $\langle (\Delta x)^2\rangle < \langle (\Delta
x)^2\rangle_0$ (resp.\ $\langle (\Delta p)^2\rangle < \langle (\Delta
p)^2\rangle_0$) or, in other words, if the ratio $X \equiv \langle (\Delta
x)^2\rangle/\langle (\Delta x)^2\rangle_0$ (resp.\ $P \equiv \langle (\Delta
p)^2\rangle/\langle (\Delta p)^2\rangle_0$) is less than one. From
(\ref{eq:dispersions}) and~(\ref{eq:dispersions-vac}), it is obvious that
the results
will be different according to whether $\lambda=2$ or $\lambda>2$.\par
%
%
{}For $\lambda = 2$, we first note that $X$ and $P$ are related with each
other by
the transformation ${\rm Re} z \to - {\rm Re} z$. Then it is clear that the
maximum
squeezing in $x$ will be achieved for real, negative values of $z$. So
hereafter we
only consider $X$ for such values. We find
\begin{equation}
  X \simeq 1 - \frac{2}{\bbeta_1} (-z) + \cdots \qquad \mbox{\rm if\ } -z \ll 1,
  \qquad X \simeq \frac{1}{2\bbeta_1} + \cdots \qquad \mbox{\rm if\ } -z \gg 1,
\end{equation}
showing that for sufficiently small values of $-z$, $X$ is always smaller
than one
and closer to zero for small values of $\bbeta_1$ than for large ones, while for
large values of $-z$, $X<1$, $X \simeq 1$, or $X>1$ according to whether
$\bbeta_1
> \frac{1}{2}$, $\bbeta_1 = \frac{1}{2}$, or $\bbeta_1 < \frac{1}{2}$. Hence, as
displayed in Fig.~3, we obtain a large squeezing effect over the whole range of
real, negative values of $z$ for positive values of $\alpha_0$, whereas for
$\alpha_0 = 0$ or $\alpha_0 < 0$, the squeezing effect becomes negligeably small
or disappears for large values of $-z$.\par
%
%
{}For $\lambda > 2$, the ratios $X$ and $P$ are equal and only depend on $|z|$.
From~(\ref{eq:dispersions}), (\ref{eq:H_0}),
and~(\ref{eq:dispersions-vac}), it is
then obvious that $X = P > 1$ if $|z| \ne 0$, so that there is no squeezing
effect in
this case.\par
%
%
It is also interesting to study higher-order squeezing~\cite{hong} in the CS
$|z;0\rangle$. The quadrature $x$ (resp.\ $p$) is said to be squeezed to
the $2N$th
order if $\langle (\Delta x)^{2N}\rangle < \langle (\Delta
x)^{2N}\rangle_0$ (resp.\
$\langle (\Delta p)^{2N}\rangle < \langle (\Delta p)^{2N}\rangle_0$).
Considering
fourth-order squeezing, we have to determine whether the ratio $Y \equiv \langle
(\Delta x)^4\rangle/\langle (\Delta x)^4\rangle_0$ (resp.\ $Q \equiv \langle
(\Delta p)^4\rangle/\langle (\Delta p)^4\rangle_0$) is less than one. For
$\mu=0$, we obtain
\begin{eqnarray}
  \left. \begin{array}{c}
           \langle (\Delta x)^4\rangle \\[0.2cm]
           \langle (\Delta p)^4\rangle
           \end{array}\right\}
           & =&  \frac{3}{2} \langle H_0^2\rangle - \frac{\lambda}{4} \left(1 +
           \bbeta_1 - \bbeta_2 - \bbeta_{\lambda-1}\right) \langle H_0\rangle +
           \frac{\lambda^2}{8} \bbeta_1 \left(1 + \bbeta_2 -
           \bbeta_{\lambda-1}\right) \nonumber \\
  &&\mbox{} + \delta_{\lambda,2} \left[z^2 + \left(z^*\right)^2 \pm 2 \left(z +
           z^*\right)(\langle H_0\rangle + 1)\right] + \delta_{\lambda,4}
\left(z +
           z^*\right), \label{eq:4order}\\
  \langle (\Delta x)^4\rangle_0 & = & \langle (\Delta p)^4\rangle_0 =
           \frac{\lambda^2}{4} \bbeta_1 (\bbeta_1 + \bbeta_2),
\end{eqnarray}
where on the right-hand side of~(\ref{eq:4order}), the upper (resp.\ lower) sign
applies to $\langle (\Delta x)^4\rangle$ (resp.\ $\langle (\Delta
p)^4\rangle$), and
\begin{equation}
  \langle H_0^2\rangle = \lambda^2 \left\{\frac{1}{4} \bbeta_1^2 + \left(
  \prod_{\nu=1}^{\lambda-1} \bbeta_{\nu}\right)^{-1} y \left[(1 + \bbeta_1 -
  \bbeta_{\lambda-1}) \Phi^{\lambda-1}_0(y) + \bbeta_{\lambda-1}
  \Phi^{\lambda-2}_0(y)\right]\right\}.
\end{equation}
\par
%
%
{}For $\lambda=2$, $Y$ and $Q$ are related with each other by the
transformation
${\rm Re} z \to - {\rm Re} z$ and the maximum fourth-order squeezing in $x$ is
achieved for real, negative values of $z$. So we only consider $Y$ for such
values
and find
\begin{equation}
  Y \simeq 1 - \frac{4}{\bbeta_1} (-z) + \cdots \qquad \mbox{\rm if\ } -z \ll 1,
  \qquad Y \simeq \frac{3}{4\bbeta_1(1+\bbeta_1)} + \cdots \qquad \mbox{\rm
  if\ } -z \gg 1,
\end{equation}
showing that the behaviour of $Y$ in terms of $-z$ and $\bbeta_1$ should be
roughly
similar to that of $X$. This is confirmed numerically, as displayed in
Fig.~3. There
are however two main differences between the results for $X$ and $Y$. First, for
$\alpha_0=0$, there is fourth-order squeezing only for $-z < 3/4$, as compared
with second-order squeezing over the whole range of $-z$ values. Second, for
$\alpha_0 > 0$ and any given $-z$ value, the fourth-order squeezing is
larger than
the second-order one.\par
%
%
{}For $\lambda>2$, we have only investigated the case $\lambda=4$, for
which it is
known that there is fourth-order squeezing when $\alpha_{\mu} = 0$~\cite{buzek}.
The present calculations show that the latter is small ($Y_{\rm min} \simeq
0.933$) and is obtained for very small values of $-z$ ($-z < 0.1$). Considering
nonvanishing values of $\alpha_{\mu}$ can enhance the fourth-order
squeezing, but
it always remains confined to rather small values of $-z$ and its
dependence on the
parameters is rather weak. For instance, we obtained fourth-order squeezing for
$-z \le 0.558$ and $Y_{\rm min} \simeq 0.632$ for $\alpha_0 = \alpha_1 = 0$ and
an $\alpha_2$ value as high as 30.\par
%
%
While retrieving the second- and fourth-order squeezing properties of standard
$\lambda$-photon CS~\cite{buzek}, we have therefore shown that they can be
improved by considering nonvanishing values of $\alpha_{\mu}$. The most striking
effect is obtained for $\lambda=2$ and $\alpha_0>0$. Note that these values of
$\alpha_0$ are precisely those for which the conventional uncertainty
constraint is
respected.\par
%
%
\section{Conclusion}

In the present Letter, we established that the $C_{\lambda}$-extended oscillator
spectrum generating algebra is a $C_{\lambda}$-extended ($\lambda-1$)th-degree
polynomial deformation of su(1,1), and we characterized its $\lambda$ unirreps
corresponding to the various subspaces ${\cal F}_{\mu}$, $\mu=0$, 1,~\ldots,
$\lambda-1$, of $\cal F$ by the eigenvalues of its Casimir operator and the
lowest
$J_0$ eigenvalue.\par
%
%
We then constructed the CS $|z; \mu\rangle$ of the spectrum generating algebra,
defined as the eigenstates of the lowering generator $J_-$. We proved that
they are
normalizable, continuous in the parameter $z$, and that they allow a
resolution of
unity. We showed that they contain as special cases both the CS of the
Calogero-Vasiliev oscillator (equivalent to some Barut-Girardello
CS~\cite{barut})
and the standard $\lambda$-photon CS of Ref.~\cite{buzek} (equivalent to some
Mittag-Leffler CS~\cite{sixdeniers}).\par
%
%
{}Finally, we established that the $C_{\lambda}$-extended oscillator parameters
have a striking influence on the CS nonclassical properties, which may be rather
different from those of standard $\lambda$-photon CS. Especially for
$\mu=0$, the
CS exhibit strong nonclassical properties, such as antibunching and quadrature
squeezing, on a considerable parameter range.\par
%
%
The CS presented here are not the only ones that can be associated with the
$C_{\lambda}$-extended oscillator. Other possibilities are under current
investigation, and we hope to report on them in the near future.\par
%
%
\section*{Acknowledgement}

The author would like to thank an anonymous referee for valuable comments and
questions.\par
%
%

%
%
\newpage
\section*{Figure captions}

{}Fig.\ 1. Mandel's parameter $Q$ as a function of $|z| \equiv r$ for
$\lambda=2$ and
various parameters: (a) $\mu=0$ and $\alpha_0 = 0$ (solid line), $\alpha_0
= - 4/5$
(dashed line), $\alpha_0 = - 24/25$ (dotted line), or $\alpha_0 = 1$ (dot-dashed
line); (b) $\mu=1$ and $\alpha_0 = 0$ (solid line), $\alpha_0 = 1$
(dot-dashed line),
$\alpha_0 = 9$ (dotted line), or $\alpha_0 = 19$ (dashed line).

\medskip\noindent
{}Fig.\ 2. Mandel's parameter $Q$ as a function of $|z| \equiv r$ for
$\lambda=3$ and
various parameters: (a) $\mu=0$ and $\alpha_0 = \alpha_1 = 0$ (solid line),
$\alpha_0 = - \alpha_1 = - 7/10$ (dashed line), $\alpha_0 = - 47/50$,
$\alpha_1 =
- 1$ (dotted line), or $\alpha_0 = - \alpha_1 = 2$ (dot-dashed line); (b)
$\mu=1$ and
$\alpha_0 = \alpha_1 = 0$ (solid line), $\alpha_0 = - \alpha_1 = 2$ (dashed
line),
$\alpha_0 = 0$, $\alpha_1 = 13$ (dotted line), or $\alpha_0 = - 47/50$,
$\alpha_1 = - 1$ (dot-dashed line); (c) $\mu=2$ and $\alpha_0 = \alpha_1 = 0$
(solid line), $\alpha_0 = - \alpha_1 = 2$ (dashed line), $\alpha_0 = 0$,
$\alpha_1 = 28$ (dotted line), or $\alpha_0 = - 47/50$, $\alpha_1 = - 1$
(dot-dashed line).

\medskip\noindent
{}Fig.\ 3. The ratios $X \equiv \langle (\Delta x)^2\rangle/\langle
(\Delta x)^2\rangle_0$ and $Y \equiv \langle (\Delta x)^4\rangle/\langle
(\Delta x)^4\rangle_0$ as functions of $-z$ for real $z$, $\lambda=2$, and
$\mu=0$. The parameter value is $\alpha_0 = 0$ (solid lines), $\alpha_0 = - 2/5$
(dashed lines), $\alpha_0 = 1$ (dotted lines), or $\alpha_0 = 3$ (dot-dashed
lines).
\par


\newpage
\begin{thebibliography}{99}

\bibitem{glauber} R.J.\ Glauber, Phys.\ Rev.\ 131 (1963) 2766.

\bibitem{perelomov} A.P.\ Perelomov, Generalized Coherent States and Their
Applications (Springer, Berlin, 1986).

\bibitem{kimble} H.J.\ Kimble, M.\ Dagenais, L.\ Mandel, Phys.\ Rev.\ Lett.\ 39
(1977) 691.

\bibitem{short} R.\ Short, L.\ Mandel, Phys.\ Rev.\ Lett.\ 51 (1983) 384;\\
M.C.\ Teich, B.E.A.\ Saleh, J.\ Opt.\ Soc.\ Am.\ B 2 (1985) 275.

\bibitem{slusher} R.E.\ Slusher,  L.W.\ Hollberg, B.\ Yurke, J.C.\ Mertz,
J.F.\ Valley,
Phys.\ Rev.\ Lett.\ 55 (1985) 2409;\\
L.-A.\ Wu, H.J.\ Kimble, J.L.\ Hall, H.\ Wu, Phys.\ Rev.\ Lett.\ 57 (1986) 2520.

\bibitem{hong} C.K.\ Hong, L.\ Mandel, Phys.\ Rev.\ Lett.\ 54 (1985) 323.

\bibitem{hillery} M.\ Hillery, Phys.\ Rev.\ A 36 (1987) 3796;\\
Y.\ Xia, G.\ Guo, Phys.\ Lett.\ A 136 (1989) 281.

\bibitem{dodonov} V.V.\ Dodonov, I.A.\ Malkin, V.I.\ Man'ko, Physica 72
(1974) 597.

\bibitem{barut} A.O.\ Barut, L.\ Girardello, Commun.\ Math.\ Phys.\ 21
(1971) 41.

\bibitem{buzek} V.\ Bu\v zek, I.\ Jex, Tran Quang, J.\ Mod.\ Opt.\ 37
(1990) 159.

\bibitem{dariano} G.M.\ D'Ariano, M.G.\ Rasetti, J.\ Katriel, A.I.\ Solomon, in:
Squeezed and Nonclassical Light, eds.\ P.\ Tombesi, E.R.\ Pike (Plenum, New
York,
1989) p.\ 301.

\bibitem{solomon} A.I.\ Solomon, Phys.\ Lett.\ A 196 (1994) 29.

\bibitem{katriel} J.\ Katriel, A.I.\ Solomon, Phys.\ Rev.\ A 49 (1994) 5149;\\
A.I.\ Solomon, in: Fifth Int.\ Conf.\ on Squeezed States and Uncertainty
Relations,
Balatonfured, Hungary, May 27--31, 1997, eds.\ D.\ Han, J.\ Jansky, Y.S.\
Kim, V.I.\
Man'ko (NASA Goddard Space Flight Center, Greenbelt, MD, 1998) p.\ 157.

\bibitem{shanta} P.\ Shanta, S.\ Chaturvedi, V.\ Srinivasan, R.\
Jagannathan, J.\
Phys.\ A 27 (1994) 6433.

\bibitem{matos} R.L.\ de Matos Filho, W.\ Vogel, Phys.\ Rev.\ A 54 (1996) 4560.

\bibitem{manko} V.I.\ Man'ko, G.\ Marmo, F.\ Zaccaria, E.C.G.\ Sudarshan, Phys.\
Scr.\ 55 (1997) 528.

\bibitem{mancini} S.\ Mancini, Phys.\ Lett.\ A 233 (1997) 291;\\
S.\ Sivakumar, Phys.\ Lett.\ A 250 (1998) 257.

\bibitem{liu} X.-M.\ Liu, J.\ Phys.\ A 32 (1999) 8685.

\bibitem{cq98} C.\ Quesne, N.\ Vansteenkiste, Phys.\ Lett.\ A 240 (1998) 21.

\bibitem{cq99} C.\ Quesne, N.\ Vansteenkiste, Helv.\ Phys.\ Acta 72 (1999) 71;
$C_{\lambda}$-extended oscillator algebras and some of their deformations and
applications to quantum mechanics, preprint math-ph/0003025, to be published in
Int.\ J.\ Theor.\ Phys.

\bibitem{klauder} J.\ Klauder, J.\ Math.\ Phys.\ 4 (1963) 1058.

\bibitem{daska} C.\ Daskaloyannis, J.\ Phys.\ A 24 (1991) L789.

\bibitem{moshinsky} M.\ Moshinsky, Yu.\ F.\ Smirnov, The Harmonic Oscillator in
Modern Physics (Harwood, Amsterdam, 1996).

\bibitem{brze} T.\ Brzezi\'nski, I.\ L.\ Egusquiza, A.\ J.\ Macfarlane,
Phys.\ Lett.\ B
311 (1993) 202.

\bibitem{chaturvedi} S.\ Chaturvedi, V.\ Srinivasan, Phys. Rev. A 44 (1991)
8024.

\bibitem{mukunda} N.\ Mukunda, E.C.G.\ Sudarshan, J.K.\ Sharma, C.L.\ Mehta, J.\
Math.\ Phys.\ 21 (1980) 2386.

\bibitem{poly} A.\ P.\ Polychronakos, Mod.\ Phys.\ Lett.\ A 5 (1990) 2325; \\
M.\ Ro\v cek, Phys.\ Lett.\ B 255 (1991) 554.

\bibitem{junker} G.\ Junker, P.\ Roy, Phys.\ Lett.\ A 257 (1999) 113;\\
D.J.\ Fern\'andez, V.\ Hussin, J.\ Phys.\ A 32 (1999) 3603.

\bibitem{erdelyi54} A.\ Erd\'elyi, W.\ Magnus, F.\ Oberhettinger, F.G.\ Tricomi,
Tables of Integral Transforms, vol.\ I (Mc-Graw Hill, New York, 1954).

\bibitem{deenen} J.\ Deenen, C.\ Quesne, J.\ Math.\ Phys.\ 25 (1984) 2354;\\
D.J.\ Rowe, J.\ Math.\ Phys.\ 25 (1984) 2662;\\
D.J.\ Rowe, G.\ Rosensteel, R.\ Gilmore, J.\ Math.\ Phys.\ 26 (1985) 2787.

\bibitem{erdelyi53} A.\ Erd\'elyi, W.\ Magnus, F.\ Oberhettinger, F.G.\ Tricomi,
Higher Transcendental Functions, vols.\ I, II, III (Mc-Graw Hill, New York,
1953).

\bibitem{sixdeniers} J.-M.\ Sixdeniers, K.A.\ Penson, A.I.\ Solomon, J.\
Phys.\ A 32
(1999) 7543.

\end {thebibliography}
\end{document}